\documentclass[conference]{IEEEtran}
\IEEEoverridecommandlockouts
\usepackage[utf8]{inputenc}
\usepackage{cite}
\usepackage{caption}
\usepackage{subcaption}
\usepackage{graphicx}
\usepackage{amsmath,amssymb,amsfonts}
\usepackage{algorithmic}
\usepackage{graphicx}
\usepackage{textcomp}
\usepackage{xcolor}
\def\BibTeX{{\rm B\kern-.05em{\sc i\kern-.025em b}\kern-.08em
    T\kern-.1667em\lower.7ex\hbox{E}\kern-.125emX}}
\begin{document}

\title{Generative AI Adoption in Classroom in Context of Technology Acceptance Model (TAM) and the Innovation Diffusion Theory (IDT)\\}

\author{\IEEEauthorblockN{Aashish Ghimire}
\IEEEauthorblockA{\textit{Department of Computer Science} \\
\textit{Utah State University}\\
Logan, Utah \\
aashish.ghimire@usu.edu}
\and
\IEEEauthorblockN{John Edwards}
\IEEEauthorblockA{\textit{Department of Computer Science} \\
\textit{Utah State University}\\
Logan, Utah \\
john.edwards@usu.edu}

}
\maketitle

\begin{abstract}
The burgeoning development of generative artificial intelligence (GenAI) and the widespread adoption of large language models (LLMs) in educational settings have sparked considerable debate regarding their efficacy and acceptability. Despite the potential benefits, the assimilation of these cutting-edge technologies among educators exhibits a broad spectrum of attitudes, from enthusiastic advocacy to profound skepticism. This study aims to dissect the underlying factors influencing educators' perceptions and acceptance of GenAI and LLMs. We conducted a survey among educators and analyzed the data through the frameworks of the Technology Acceptance Model (TAM) and Innovation Diffusion Theory (IDT). Our investigation reveals a strong positive correlation between the perceived usefulness of GenAI tools and their acceptance, underscoring the importance of demonstrating tangible benefits to educators. Additionally, the perceived ease of use emerged as a significant factor, though to a lesser extent, influencing acceptance. Our findings also show that the knowledge and acceptance of these tools is not uniform, suggesting that targeted strategies are required to address the specific needs and concerns of each adopter category to facilitate broader integration of AI tools in education.
\end{abstract}


\section{Introduction}
The advent of generative artificial intelligence (GenAI) has heralded a new era in the technological landscape, offering unprecedented capabilities in creating text, images, code, and more from simple prompts. Among its various applications, the potential use of GenAI in educational settings is particularly compelling. Large language models (LLMs), a subset of GenAI, are poised to revolutionize teaching and learning practices by providing personalized learning experiences, automating content generation, and facilitating a more interactive and engaging learning environment. These technologies can augment the educational process, from crafting tailored educational materials to supporting diverse learning strategies, thereby enhancing the efficacy and accessibility of education. Furthermore, GenAI's ability to analyze and generate complex data can significantly contribute to research methodologies, enabling educators and students alike to explore new frontiers of knowledge and learning.

However, the integration of GenAI and LLMs into classroom settings is not without challenges. The adoption of new technologies in education is influenced by a multitude of factors, including but not limited to, perceived usefulness, ease of use, and the technological infrastructure available. To understand the dynamics of these technologies' acceptance and integration, it is crucial to delve into established theoretical frameworks that explain the adoption of technological innovations. The Technology Acceptance Model (TAM) and Innovation Diffusion Theory (IDT) offer robust lenses through which to examine these phenomena.

The Technology Acceptance Model (TAM)\cite{davis1987user} posits that the perceived usefulness and perceived ease of use are fundamental determinants of the acceptance and usage of new technology. According to TAM, if users believe a technology will enhance their job performance (usefulness) and will be free of effort (ease of use), they are more likely to embrace and utilize the technology. On the other hand, Innovation Diffusion Theory (IDT) proposed by Rogers, explores how, why, and at what rate new ideas and technology spread through cultures~\cite{rogers1961bibliography,rogers1983diffusion}. IDT suggests that innovation adoption is influenced by factors such as the innovation's relative advantage, compatibility with existing values and practices, complexity or ease of use, trialability, and observable results. Together, these frameworks provide a comprehensive understanding of the multifaceted process of technological adoption, enabling a nuanced analysis of the barriers and drivers behind GenAI's integration into the educational sphere. 
This paper has one primary research question: 
\begin{itemize}
    \item[RQ] \textbf{What facilitators and barriers to the adoption of generative AI technologies exist in educational settings?}
\end{itemize}

In this paper, we aim to answer our research question by examining educators' perceptions and acceptance of GenAI and LLMs through the TAM and IDT frameworks. Understanding these factors is crucial for developing strategies to encourage the effective integration of GenAI tools in classrooms, thereby maximizing their potential benefits for teaching and learning. The following sections will delve into the methodology of our study, present our findings, and discuss their implications for the future of GenAI in education, setting the context for a comprehensive exploration of GenAI's role in reshaping educational paradigms. This inquiry not only contributes to the academic discourse on educational technology adoption but also provides practical insights for educators, policymakers, and technology developers aiming to foster an environment conducive to the innovative use of GenAI in education.

\section{Related works}
\subsection{Teachers' perspectives on AI in education}
Understanding the attitudes and perceptions of educators towards AI in education is crucial for its acceptance and integration into teaching practices. A survey of Kenyan teachers by Bii et al. revealed a generally positive outlook towards chatbot usage in education, despite concerns regarding their accuracy and potential to replace human teachers~\cite{bii2018teacher}. Similarly, Zhai et al.'s content analysis highlighted key research areas in AI education over a decade, including development and application~\cite{Zhai2021Review}. Chen et al. noted an increased academic focus on AI, particularly in natural language processing and neural networks for educational purposes~\cite{chen2022two}.

Research by Guillén-Gámez and Mayorga-Fernández found that factors such as age, gender, and ICT project involvement positively influence educators' attitudes towards ICT use in higher education~\cite{Guillén-Gámez_Mayorga-Fernández_2020}. Conversely, Nazaretsky et al. identified confirmation bias and trust as significant influencers on teachers’ attitudes towards AI-based technologies, suggesting that pre-existing beliefs could hinder the adoption of such tools~\cite{Nazaretsky_Cukurova_Ariely_Alexandron_2021}.

Akgun and Greenhow emphasized the ethical considerations necessary for AI deployment in K-12 settings, advocating for principles like transparency and inclusiveness~\cite{Akgun_Greenhow_2022}. Celik et al. explored the multifaceted roles of teachers in AI research and the challenges faced, including technical limitations and lack of technological knowledge~\cite{Celik_Dindar_Muukkonen_Järvelä_2022}. Kim and Kim's study on STEM teachers' perceptions of an AI-enhanced scaffolding system for scientific writing indicated positive expectations, yet highlighted the need for teacher training on AI technologies~\cite{Kim_Kim_2022}. Lastly, Lau and Guo's investigation into university instructors' views on AI tools like ChatGPT in programming education uncovered diverse strategies for adaptation, raising important questions for future research in computing education~\cite{lau2023ban}.

\subsection{ Technology Acceptance Model (TAM) and Innovation Diffusion Theory (IDT) to Explore the Adoption of Technology}

The TAM, proposed by Fred Davis in 1989, posits that Perceived Usefulness (PU) and Perceived Ease of Use (PEOU) play a critical role in user acceptance of information systems~\cite{davis1987user}. Masrom investigated the learning acceptance in terms of TAM and found that TAM could largely explain it's acceptance~\cite{masrom2007technology}. L Ritter  performed a meta-analysis employing meta-analytic structural equation modeling (MASEM) to quantitatively synthesize studies that investigates college students’ acceptance of online learning managements systems and got mixed results on how well it fits TAM\cite{l2017technology}.  Scherer et al. performed a meta-analytic structural equation modeling to investigate the Technology Acceptance Model's (TAM) validity in explaining teachers' adoption of digital technology in education~\cite{Scherer_Siddiq_Tondeur_2019}. Through robust statistical techniques, the study provided a comprehensive understanding of the factors influencing teachers' acceptance and use of technology, highlighting the role of perceived usefulness and ease of use. The results demonstrated the strong predictive power of TAM in teachers' technology adoption, offering a valuable framework for future research and technology integration strategies in the educational context. The role of certain key constructs and the importance of external variables contrast some existing beliefs about the TAM. Granic and  Marangunic in their meta studey of 71 related papers  found that TAM and its many different versions represent a credible model for facilitating assessment of diverse learning technologies and TAM's core variables, perceived ease of use and perceived usefulness, have been proven to be antecedent factors affecting acceptance of learning with technology. Zaineldeen et. al studied the TAM's concepts, contribution, limitation, and adoption in education~\cite{zaineldeen2020technology}. 

Chocarro et al.'s application of the Technology Acceptance Model (TAM) to teachers' attitudes towards chatbots showed a preference for formal language and indicated that age and digital skills play roles in acceptance~\cite{Chocarro_Cortiñas_Marcos-Matás_2023}. Khong et al. extended TAM to understand factors affecting teachers’ acceptance of technology for online teaching, finding cognitive attitudes and perceived usefulness to be significant predictors~\cite{Khong_Celik_Le_Lai_Nguyen_Bui_2023}. A 2023 study by Iqbal et al. on faculty attitudes towards ChatGPT using TAM revealed mixed perceptions, with concerns about cheating balanced against the tool's benefits for lesson planning~\cite{Iqbal_Ahmed_Azhar_2023}. 

Similarly, innovation diffusion theory (IDT) have been used to study the acceptance and spreading of technology in education. Pinho et al.'s study on Moodle's use in higher education identified positive influences of Moodle's characteristics and personal innovativeness on its adoption, highlighting the importance of student-centered Learning Management Systems (LMS) ~\cite{pinho2021application}. Sahin provides a comprehensive overview of Rogers' Diffusion of Innovations theory, elaborating on its four main elements, the innovation-decision process, attributes of innovations, adopter categories, and its application in educational technology studies ~\cite{sahin2006detailed}. Menzli et al. examined the adoption of Open Educational Resources (OER) in higher education, finding that attributes such as relative advantage and observability positively impact faculty adoption, while also emphasizing the role of trialability, complexity, and compatibility in increasing OER adoption rates ~\cite{menzli2022investigation}. Frei-Landau et al. explored the mobile learning (ML) adoption process among teachers during the COVID-19 pandemic, uncovering 12 themes that denote the ML adoption process through Rogers' IDT, providing insights into promoting ML in teacher education under both routine and emergency conditions ~\cite{frei2022using}. Finally, Al-Rahmi et al. combined the Technology Acceptance Model (TAM) with IDT to investigate students' intentions to use e-learning systems, demonstrating that innovation characteristics significantly influence students' behavioral intentions towards e-learning systems ~\cite{al2019integrating}. Ghimire et. al. explored the students'~\cite{ghimire2024coding}, educators'~\cite{ghimire2024generativeEducators}, and administrators'~\cite{ghimire2024guidelines} attitude towards these generative AI based tools and found them to be generally positive while also highlighting various challenges~\cite{ghimire2024generative}.

\section{Methodology - Evaluation Framework}

\subsection{Survey and Data}

We conducted a quantitative study using a survey to gather educators' perspectives on AI tools in the classroom. We distributed the survey via email to faculty members at Utah State University (USU), a mid-sized research university in the western United States. Each faculty member received the survey link only once to avoid duplicate responses. The email provided a brief introduction to the research study, assured confidentiality, and encouraged participation. Participants were informed about the voluntary nature of the survey. 

We received a total of 116 survey responses from email requests, representing a diverse sample from 8 colleges and 23 out of 39 departments at the university. The wide-ranging representation ensures a comprehensive understanding of educators' attitudes from various academic disciplines. For this study, we selected six survey questions that directly support our analysis using the Technology Acceptance Model (TAM) and Innovation Diffusion Theory (IDT) frameworks. Responses were captured using a Likert scale, allowing participants to express their agreement or disagreement with specific statements. The survey approved by the USU ethics review board (IRB). Since TAM identifies Perceived Usefulness (PU) and Perceived Ease of Use (PEOU) as key determinants of technology adoption, the following questions were designed to represent these constructs:

\begin{enumerate}
\item AI tools like ChatGPT and Bard should be allowed and integrated into education.  --- \textbf{This question measures acceptance of technology, aligning with TAM's focus on the behavioral intention to use technology.}
\item I believe that AI tools like ChatGPT and Bard enhance the quality of education. --- (\textbf{Q\textsuperscript{PU}})
\item I believe the benefits of incorporating large language models in education outweigh the potential risks and ethical concerns. --- (\textbf{Q\textsuperscript{PU}})
\item I believe that the tools like ChatGPT and Bard are easy to use. ---- (\textbf{Q\textsuperscript {PEOU}})
\item I believe that these AI tools like ChatGPT and Bard could be easily integrated into my current teaching methodology. --- (\textbf{Q\textsuperscript{PEOU}})
\item Are you familiar with AI tool such as ChatGPT or Google Bard? -- (\textbf{QIDT\textsuperscript{FM}})
\end{enumerate}

Questions tagged with (\textbf{Q\textsuperscript{PU}}) measure Perceived Usefulness (PU), and those with (\textbf{Q\textsuperscript{PEOU}}) assess Perceived Ease of Use (PEOU). The question marked (\textbf{QIDT\textsuperscript{FM}}) gauges familiarity, an important aspect of IDT.

\subsection{Technology Acceptance Model (TAM) as an Evaluation Framework}

Applying the Technology Acceptance Model (TAM) to understand teachers' attitudes and perceptions towards AI tools and Large Language Models (LLMs) such as ChatGPT and Bard can offer valuable insights. In this study's context, PU encompasses teachers' belief that specific AI tools or LLMs will enhance their teaching effectiveness and student learning outcomes. Conversely, PEOU refers to the ease with which educators can utilize these tools. Factors influencing PEOU include the user interface design, learning curve, and availability of technical support, which can significantly impact teachers' willingness to adopt AI technologies. Additionally, TAM helps identify potential barriers to technology adoption, such as perceived lack of IT skills or negative attitudes towards technology, guiding the development of professional training programs to mitigate these challenges and promote positive engagement with AI and LLMs in educational settings.

\subsection{The Innovation Diffusion Theory (IDT) as an Evaluation Framework}

The Innovation Diffusion Theory (IDT), proposed by Everett Rogers in 1962, offers a comprehensive framework for understanding the mechanisms through which new ideas and technologies are adopted within social systems. IDT delineates four key elements that influence the dissemination of an innovation: the characteristics of the innovation itself, the communication channels used to spread information about the innovation, the passage of time, and the nature of the social system. The theory categorizes the adoption process into five sequential stages:

\begin{enumerate}
    \item Knowledge: This initial phase involves becoming aware of the innovation, albeit without detailed information about its functionality or application.
    \item Persuasion: At this stage, interest in the innovation grows, prompting an active search for more information and a better understanding of its benefits and drawbacks.
    \item Decision: Here, individuals or organizations critically assess the innovation, considering the pros and cons before making a decision to adopt or reject it.
    \item Implementation: During implementation, the innovation is actively integrated into use, with adjustments and adaptations often made to fit specific needs.   
    \item Confirmation: In this final stage, the effectiveness and utility of the innovation are evaluated, influencing the decision to continue its use based on observed outcomes.
\end{enumerate}

Moreover, IDT classifies adopters into five groups according to their propensity to embrace new technologies: Innovators, Early Adopters, Early Majority, Late Majority, and Laggards. This categorization helps in understanding the adoption timeline within a social system.


\section{Results}
\subsection{Using TAM as a Framework}
\begin{figure}[h]
    \centering
    \includegraphics[width=0.49\textwidth]{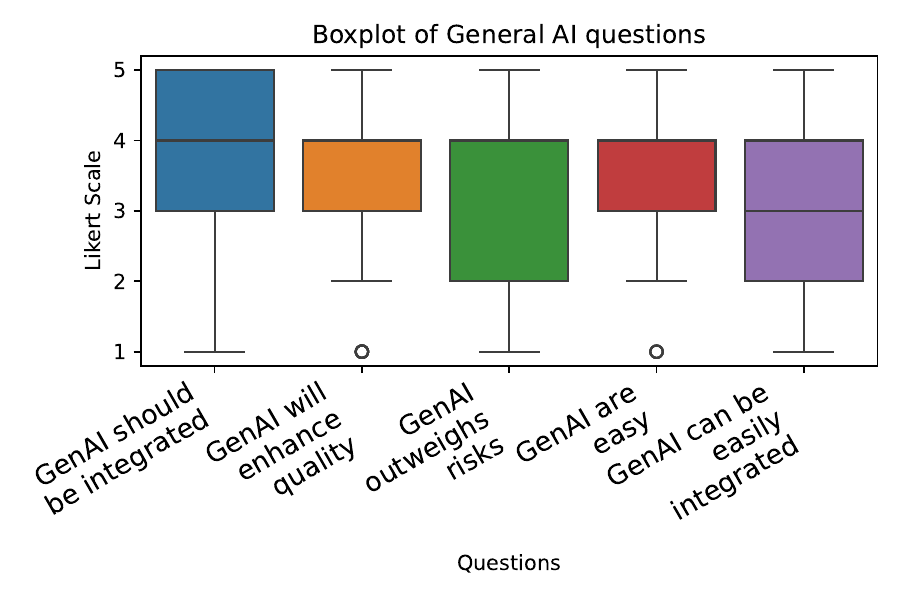}
    \caption{Raw answers to the TAM-related questions}
    \label{fig:tam-box-raw}
\end{figure}

As explained in the methodology section, we utilized five survey questions to align with the TAM framework. Since the responses were on a Likert scale, they could be directly converted to numeric values. The response to the statement \textit{``AI tools like ChatGPT and Bard should be allowed and integrated into education''} serves as a direct substitute for the dependent variable `acceptance', as integrating it into coursework signifies full acceptance of the tool. For perceived usefulness (PU), we averaged the responses to the statements \textit{``I believe that AI tools like ChatGPT and Bard enhance the quality of education''} and \textit{``I believe the benefits of incorporating large language models in education outweigh the potential risks and ethical concerns''}. For perceived ease of use (PEOU), we averaged the responses to the statements \textit{``I believe that the tools like ChatGPT and Bard are easy to use''} and \textit{``I believe that these AI tools like ChatGPT and Bard could be easily integrated into my current teaching methodology.''} This approach was adopted because the ease of use by educators should not only consider their own ease of use but also the ease of integrating it into their courses. Figure \ref{fig:tam-box-raw} shows the numeric Likert scale responses to these statements.

Next, we examine the correlation between acceptance, PU, and PEOU using the Pearson correlation coefficient.

As shown in Table \ref{tab:corr-table}, a strong positive correlation ($r = 0.734$) was found, indicating that as perceived usefulness increases, acceptance also tends to increase. A moderate positive correlation ($r = 0.542$) between acceptance and perceived ease of use was also observed.

\begin{table}[h]
    \centering
    \begin{tabular}{c c c}
    \hline
         & Perceived Usefulness & Perceived Ease of Use \\
        \hline
        Corr. with Acceptance & 0.734 & 0.542 \\
        p-value & $3.57 ^{-20}$ & $8.33 ^ {-10}$ \\
        \hline
    \end{tabular}
    \caption{Correlation table with acceptance}
    \label{tab:corr-table}
\end{table}

Regression analysis was performed to quantify how well acceptance is explained by perceived ease of use and perceived usefulness, including the significance of these predictors. It yielded an R-squared value of 0.566, indicating a moderate to strong fit. This suggests that perceived ease of use and perceived usefulness together explain a significant portion of the variance in acceptance. The coefficient for perceived usefulness was 0.678 with a p-value of $7.2^{-13}$, showing a highly significant and strong positive effect on acceptance. Perceived ease of use had a coefficient of 0.227 with a p-value of 0.026, indicating a statistically significant positive effect on acceptance. This confirms that perceived usefulness is a significant and strong predictor of acceptance. The overall model is statistically significant, as indicated by an F-statistic p-value of $4.23 \times 10^{-20}$, meaning that the predictors together significantly explain the variability in acceptance.

\begin{table}[h]
    \centering
    \begin{tabular}{c c c}
    \hline
         & Coefficient & p-value \\
        \hline
        Perceived Usefulness & 0.678 & $7.2^{-13}$ \\
        Perceived Ease of Use & 0.227 & 0.026 \\
        \hline
    \end{tabular}
    \caption{Regression analysis results}
    \label{tab:regression-analysis}
\end{table}

\subsection{The Innovation Diffusion Theory (IDT) to Explain the GenAI Use in Classrooms}

The Innovation Diffusion Theory (IDT) offers a comprehensive framework for understanding the factors that facilitate the adoption of new technological ideas or systems within society. Unlike the Technology Acceptance Model (TAM), which provides a quantitative and concise explanation of innovation adoption, IDT offers insights into the adoption phase an individual or group might be in. IDT categorizes the population into five segments based on their adoption behavior:

\begin{enumerate}
    \item Innovators: Individuals who embrace risks and are the first to experiment with new ideas.
    \item Early Adopters: Those keen on exploring new technologies and affirming their usefulness within the community.
    \item Early Majority: Individuals who contribute to mainstreaming an innovation within society, representing a significant portion of the population.
    \item Late Majority: People who adopt an innovation following its acceptance by the early majority, integrating it into their daily lives as part of the wider community.
    \item Laggards: Individuals who are slow to adopt innovative products and ideas, trailing behind the broader societal adoption curve.
\end{enumerate}

\begin{figure}[h]
    \centering
    \includegraphics[width=0.49\textwidth]{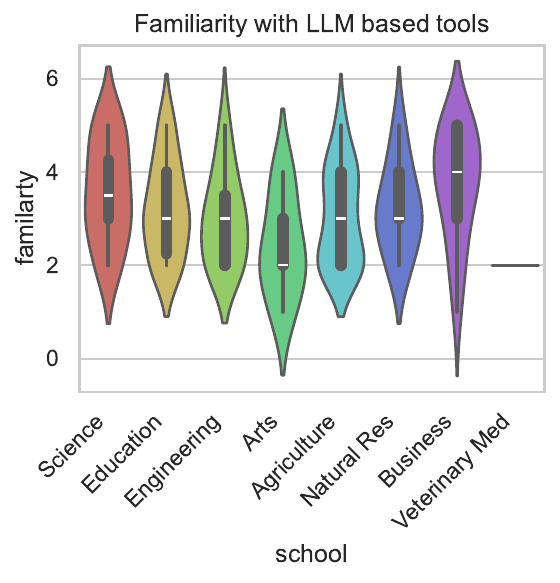}
    \caption{Violin Plot showing familiarity with LLM-based tools among educators in various colleges.}
    \label{fig:familiarity-by-school}
\end{figure}

While it is challenging to clearly categorize educators into these groups, such distinctions do exist. The range of familiarity with GenAI and LLM-based tools varies significantly across different departments and colleges. Figure \ref{fig:familiarity-by-school} shows the familiarity with these tools in various schools.

In the context of education, particularly concerning teachers' attitudes and perceptions towards AI tools and Large Language Models (LLMs) like ChatGPT and Bard, IDT provides valuable insights:

\begin{enumerate}
    \item Knowledge: Assessing teachers' awareness of AI tools and LLMs is crucial. Initiatives such as awareness campaigns, professional development sessions, and targeted marketing can significantly enhance this knowledge base.
    \item Persuasion: Understanding teachers' interest in and attitudes towards these technologies is essential. Factors like perceived usefulness and ease of use play a critical role in shaping these attitudes.
    \item Decision: The choice by teachers to incorporate AI tools into their classrooms is influenced by both individual preferences and institutional support structures.
    \item Implementation: Effective integration of AI tools into teaching practices requires adequate support, training, and resources to ensure success.
    \item Confirmation: Teachers' decisions to persist with the use of AI tools are influenced by the tangible benefits observed, feedback from students, and the availability of ongoing support.
\end{enumerate}

By identifying where teachers stand in the diffusion process and recognizing their adopter category, strategies can be customized to facilitate the adoption of AI technologies. For example, while Innovators and Early Adopters may readily experiment with new tools, the Late Majority and Laggards might need more substantial evidence of the tools' effectiveness and comprehensive support systems to be persuaded.

ChatGPT became the fastest technology product to ever reach 100 million active users~\cite{reuters2023}. The spread of the technology is so rapid that it is challenging to gauge the sense of spread or adoption in the general public. Even in education, AI tools like these are rapidly becoming commonplace. Among the five steps of innovation diffusion outlined by IDT - knowledge, persuasion, decision, implementation, and confirmation - we could use the survey responses as proxies for some of the steps. For example, the knowledge step can be directly analogous to the question asking about familiarity with the AI tools. Similarly, the implementation and confirmation could be the execution of integrating the AI tool in class and its result, which are out of the scope of this paper.

\section{Discussion and Conclusions}

This paper explored the adoption and integration of generative artificial intelligence (GenAI) and large language models (LLMs) in educational settings, using the Technology Acceptance Model (TAM) and the Innovation Diffusion Theory (IDT) as theoretical frameworks. Our survey, conducted among educators at a medium-sized public research university in the United States, provided insights into their attitudes towards the use of AI tools like ChatGPT and Bard in the classroom. The findings indicate a generally positive perception towards these technologies, underscored by the perceived usefulness (PU) and perceived ease of use (PEOU) as significant predictors of their acceptance and integration into teaching methodologies.

The analysis revealed a strong positive correlation between the perceived usefulness of AI tools and their acceptance among educators, emphasizing the importance of demonstrating tangible benefits to enhance the adoption rate. Similarly, the perceived ease of use was found to have a significant, albeit moderate, positive effect on acceptance, highlighting the need for user-friendly and accessible AI tools in educational environments. TAM is a well-established theory that has been used to study the acceptance of new technologies in a variety of contexts. However, it is important to note that TAM is not a perfect theory. It has been criticized for being too simplistic and for not taking into account the full range of factors that influence users' intention to use a technology. TAM does not take into account the full range of factors that may influence teachers' attitudes and perceptions towards these technologies, such as their beliefs about the potential benefits and risks of AI, their level of comfort with technology, and their personal experiences with AI. The model does not account for the social and cultural factors that may influence teachers’ acceptance of these tools. 

Applying IDT, we categorized educators based on their adoption behavior and identified varied levels of familiarity with GenAI and LLMs across different departments. This diversity suggests the necessity for targeted strategies to address the specific needs and concerns of each adopter category, from Innovators to Laggards, to facilitate broader and more effective integration of AI tools in education. Based on interviews with educators, as detailed separately in \cite{ghimire2024generative}, it was noted that early adopters are actively employing and incorporating these AI tools in their classes, expressing a need for clear policy guidelines. Meanwhile, laggards require training and education on the operation, advantages, and disadvantages of these tools, with many needing a combination of both approaches.

The rapid advancement of GenAI and LLMs presents a transformative opportunity for education. By embracing these technologies, educators can enhance the quality of education and foster a more engaging and personalized learning experience. Nevertheless, the successful integration of AI tools in education requires not only technological innovation but also a comprehensive understanding of the human factors influencing their adoption. Future research should therefore focus on longitudinal studies to track the evolution of educators' attitudes and the impact of AI tools on educational outcomes, as well as on the development of frameworks to address the ethical implications of AI in education.

\subsection{Threats to validity}
Our survey was not validated and no evaluation of reliability was made. Furthermore, all respondents were from a single institution based in the United States, limiting external validity regionally and internationally. Finally, generative AI is a fast-moving technology and attitudes and policies are likely also changing quickly. This work represents a shapshot of opinions, states and attitudes between May and June 2023.

\section{Future Work}

Future research should aim to extend the findings of this study by examining the long-term impact of GenAI and LLMs on educational outcomes and student engagement. Investigating the evolving attitudes of educators as they gain more experience with these technologies will also provide deeper insights into the barriers and facilitators of AI tool integration in educational settings. Additionally, exploring the ethical considerations and potential biases in AI applications in education will be crucial to ensure equitable and inclusive learning environments.

\bibliographystyle{IEEEtran}
\bibliography{references}

\end{document}